\documentclass[12pt]{iopart}
\usepackage{bm}
\usepackage{graphicx}
\usepackage{color}

\begin{document}

\title{Fluorite-related iridate Pr$_3$IrO$_7$: Crystal growth, structure, magnetism, thermodynamic, and optical properties}

\author{Harish Kumar,$^1$ M. K\"opf,$^1$ A. Ullrich,$^2$ M. Klinger,$^3$ A. Jesche$^3$ and C. A. Kuntscher$^1$}

\address{$^1$ Experimentalphysik II, Institute of Physics, Augsburg University, 86159 Augsburg, Germany}
\address{$^2$ Experimentalphysik IV, Institute of Physics, Augsburg University, 86159 Augsburg, Germany}
\address{$^3$ Experimentalphysik VI, Center of Electronic Correlations and Magnetism, Augsburg University, 86159 Augsburg, Germany}

\eads{\mailto{harish.kumar@physik.uni-augsburg.de christine.kuntscher@physik.uni-augsburg.de}}

\begin{abstract}
Spin-orbit coupling in heavy 5$d$ metal oxides, in particular, iridates have received tremendous interest in recent years due to the realization of exotic electronic and magnetic phases. Here, we report the synthesis, structural, magnetic, thermodynamic, and optical properties of the ternary iridate Pr$_3$IrO$_7$. Single crystals of Pr$_3$IrO$_7$ have been grown by the KF flux method. Structural analysis shows that Pr$_3$IrO$_7$ crystallizes in an orthorhombic phase with $Cmcm$ symmetry. The electron energy loss spectroscopy study indicates that Pr is in a 3+ valence state, which implies a 5+ oxidation state of Ir. Magnetization data measured at high and low magnetic fields do not exhibit any bifurcation between $M_{ZFC}$ and $M_{FC}$, however, a weak hump in $M(T)$ is observed at $T^*$$\sim$10.4~K. The specific heat data reveal two maxima at $\sim$253 K and $\sim$4.8 K. The optical conductivity $\sigma_1(\omega)$ spectrum shows 24 infrared-active phonon modes and reveals an insulating behavior with an optical gap $\Delta_{OP}$ of size $\sim$500~meV.
During cooling down, the temperature-dependent reflectivity spectrum reveals eight extra phonon modes below the structural phase transition ($\sim$ 253 K). An anomaly is observed at around $T^*$ in the temperature evolution of infrared-active mode frequencies suggesting the presence of significant spin-phonon coupling in the system.
\end{abstract}


\submitto{\JPCM}

\maketitle
\section {Introduction}
In recent years, iridates have drawn much attention due to a variety of exotic properties arising from the presence of spin-orbit coupling (SOC), on-site Coulomb ($U$), and crystal field ($\Delta_{CFE}$) interactions of comparable strength \cite{William.2014,Jeffrey.2015}. As compared to the 3$d$ transition metal oxides, iridates exhibit a large SOC and smaller $U$ due to the heavy and extended nature of the 5$d$ orbitals, respectively. In iridates, the comparable energy strength of SOC, $U$, and $\Delta_{CFE}$ creates a fragile balance between interactions that are predicted to host the novel topological phases i.e., Mott insulator, Weyl semimetal and axion insulator \cite{Pesin.2010, Wan.2011, Ueda.2018,Ueda.2020}. Experimentally and theoretically, these interactions have been extensively studied in the context of tetravalent Ir$^{4+}$ (5$d^5$) within the single particle scenario, where the strong SOC splits the Ir $t_{2g}$ states into a completely filled $J_{eff}$ = 3/2 band and a half-filled $J_{eff}$ = 1/2 band, which lies close to the Fermi energy. This half-filled $J_{eff}$ = 1/2 band further splits into a lower (filled) and an upper (empty) Hubbard band due to the presence of small $U$, thus leading to a $J_{eff}$ = 1/2 Mott state \cite{Kim.2008,Kim.2009,Moon.2008}. In contrast, the Ir$^{5+}$ ions have four 5$d$ electrons, the $J_{eff}$ = 3/2 band is expected to be fully filled and the $J_{eff}$ = 1/2 band to be empty, leading to a nonmagnetic $J_{eff}$ = 0 state in the strong SOC limit. However, recent theoretical and experimental studies have shown strong evidence against the SOC driven $J_{eff}$ = 0 state \cite{Kishor.2020,shobha.2021,Charu.2021,Dey.2017,Abhishek.2016,Cao.2014}. The magnetism in the Ir$^{5+}$ $J_{eff}$ = 0 picture under strong SOC is thus debatable.

Generally, the 5$d$ Ir based ternary oxide materials with chemical composition $A_3$IrO$_7$ ($A$ = rare-earth metal) belong to the family of pentavalent (Ir$^{5+}$) iridates. These materials basically belong to the class of the defect-fluorite structure with orthorhombic $Cmcm$ symmetry in which the Ir$^{5+}$ ions form IrO$_6$ octahedra and are arranged in one-dimensional chains along the $c$ axis \cite{Vente.2004,Nishimine.2004,Nishimine.2007}. Previously, the structural, magnetic and thermal properties for $A_3$IrO$_7$ ($A$ = Pr, Nd, Sm, and Eu) compounds have been studied in the polycrystalline form, in addition, a single crystal study of Pr$_3$IrO$_7$ is reported for the electrocatalyst properties \cite{Nishimine.2004,Nishimine.2007,Wang.2021}. The specific heat data for (Pr, Nd, Sm and Eu)$_3$IrO$_7$ exhibit a thermal anomaly at 261, 342, 420, 485 K, respectively, which has been later assigned to a structural phase transition \cite{Nishimine.2004,Nishimine.2007}. It was reported that among these materials only Nd$_3$IrO$_7$ reveals an antiferromagnetic (AFM) transition at around 2.6~K \cite{Nishimine.2004}.

In the present work, we report the synthesis and characterization of Pr$_3$IrO$_7$ single crystals. Molten KF flux method has been used to grow the Pr$_3$IrO$_7$ single crystals, which were characterized by Laue x-ray diffraction, electron energy loss spectroscopy, magnetization, specific heat, and infrared reflectivity measurements. The obtained magnetization data show a weak hump at $T^*$$\sim$10.4~K and a continuous rise below 5.6~K with lowering the temperature. The $M(T)$ data measured at low and high fields do not exhibit any bifurcation between $M_{ZFC}$ and $M_{FC}$ either. Our specific heat results exhibit two maxima at $\sim$253 K and $\sim$4.8 K. The infrared reflectivity measurements show several infrared-active phonon modes, and the low reflectivity value in the low-frequency range implies an insulating behavior for Pr$_3$IrO$_7$. The temperature evolution of the phonon modes clearly reveal a change across the structural phase transition  and a hump around $T^*$.

\section {Experimental Details}
High-quality ingredient powder materials Pr$_6$O$_{11}$ and IrO$_2$ with a phase purity of more than 99.99\% (M/s Sigma-Aldrich) were mixed in their stoichiometric ratios with a slight excess of IrO$_2$ and ground well. A pre-heat treatment was performed for the starting material Pr$_6$O$_{11}$ at 800$^{\circ}$C for $\sim$ 8 hours to remove any residual atmospheric moisture. The powder mixture was subsequently pelletized and sintered in air at temperatures 950 - 1130$^{\circ}$C for several days with intermediate grindings. Single crystals of Pr$_3$IrO$_7$ were grown by KF flux method using polycrystalline sample (slight excess of IrO$_2$) and KF flux in a ratio of 1:200. For the crystal growth, the sample was heated at 1100$^{\circ}$C for 4 hours and slowly cooled to 850$^{\circ}$C at a rate of 2$^{\circ}$C/hour. Many small crystals were found at the wall of the Pt-crucible [left inset of Fig.\ 1(a)]. The crystals were removed from the residual KF by rinsing them out with water. The structure and chemical composition of Pr$_3$IrO$_7$ samples were analysed using powder x-ray diffraction (XRD) and energy dispersive analysis of x-ray (EDX) using a ZEISS Crossbeam 550/550L scanning electron microscope with an Oxford detector. To check the purity of the crystals, powder XRD was performed on the powder of the crystals using a Rigaku diffractometer (Model: Miniflex 600) with Cu$K_\alpha$ radiation in 2$\theta$ range 10$^{\circ}$ to 90$^{\circ}$ with a step of 0.02$^{\circ}$. The XRD data were analysed by Rietveld refinement using the FULLPROF program. To further check the purity of the crystals, x-ray Laue diffraction experiments were performed, confirming the high crystallinity of the crystals.

The transmission electron microscopy including electron energy-loss spectroscopy (EELS) was carried out with a JEOL NEOARM F200 instrument, equipped with a Gatan OneView camera and a Continuum S EELS spectrometer. The EELS analysis was done at 200 keV beam energy in the DualEELS\textsuperscript{TM} mode. DC magnetization data were collected from 1.8~K to 300~K using a superconducting quantum interference device (SQUID, Quantum Design) magnetometer MPMS. Specific heat data were measured in the temperature range from 1.8~K to 300~K using Quantum Design PPMS DynaCool with a heat pulse relaxation method. The temperature-dependent reflectance measurements were carried out in the temperature range 300 - 5.4 K and frequency range 126 - 20000~cm$^{-1}$ using a CryoVac Konti cryostat attached to a Bruker Vertex v80 Fourier transform infrared spectrometer and an infrared microscope (Bruker Hyperion). A thin silver layer was evaporated on half of the crystal, which served as reference to obtain the absolute reflectivity.
The frequency resolution was set to 2 cm$^{-1}$ in the far-infrared range, while a resolution of 4 cm$^{-1}$  was chosen for the higher-frequency range.
For the analysis, the measured reflectivity spectrum was extrapolated in the low- and high-energy regime by Lorentz fitting and x-ray optic volumetric data, respectively \cite{Tanner.2015}. Note that a power-law $\omega^n$ interpolation with an integer $n$ = 3 was used to cover the region between measured reflectivity and calculated high-energy extrapolated data. The so-obtained reflectivity spectrum was used to obtain the real part of the optical conductivity $\sigma _1(\omega$) and the loss function  -Im(1/$\epsilon$)
using the Kramers-Kronig relations and a Lorentz model was used to fit the optical spectra using the RefFIT program \cite{Kuzmenko.2005}.

\begin{figure}
	\centering
		\includegraphics[width=9cm]{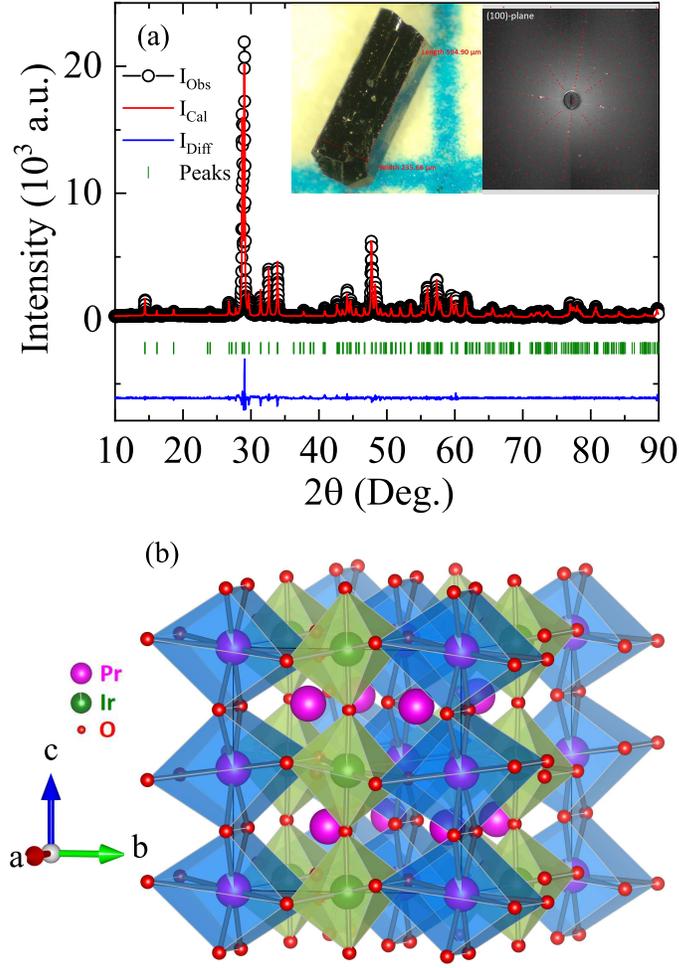}
	\caption{(a) Room-temperature XRD pattern along with Rietveld refinement for Pr$_3$IrO$_7$. Right and left inset show crystal image and x-ray Laue diffraction pattern of the Pr$_3$IrO$_7$ single crystal, respectively. (b) Crystal structure of Pr$_3$IrO$_7$ with the Pr(1)O$_8$ pseudocubes (blue) and IrO$_6$ octahedra (green).}\label{fig.XRD}
\end{figure}
\section{Results and Discussions}

\subsection {Structural Analysis}
The room-temperature XRD pattern of Pr$_3$IrO$_7$ together with the Rietveld refinement is shown in Fig.\ \ref{fig.XRD} (a). The refinement analysis confirms that Pr$_3$IrO$_7$ is in a single phase and crystallises in an orthorhombic phase with $Cmcm$ symmetry (space group 63). The unit cell parameters $a$, $b$ and $c$ are found to be 10.97 {\AA}, 7.42 {\AA} and 7.53 {\AA}, resp., which match the reported values \cite{Nishimine.2004}. We have confirmed the crystallographic planes for Pr$_3$IrO$_7$ single crystal through Laue diffraction, which insures that Pr$_3$IrO$_7$ crystals are in a unique crystalline domain [right inset of Fig.\ \ref{fig.XRD}(a)]. Fig.\ \ref{fig.XRD} (b) shows the schematic representation of the crystal structure of Pr$_3$IrO$_7$, where the Ir cations are six-fold coordinated and form one-dimensional chains of the corner-sharing IrO$_6$ octahedra along the $c$ axis. The Pr cations are eight-fold coordinated and form a chain of eightfold-distorted Pr(1)O$_8$ cubes, which simultaneously share the edges with the chains of IrO$_6$ octahedra. The remaining Pr(2) cations lie in between the Pr(1)O$_8$ pseudocubes and IrO$_6$ octahedra \cite{Nishimine.2004}.

\subsection {Electron Energy Loss Spectroscopy}
\begin{figure}
	\centering
	\includegraphics[width=8cm]{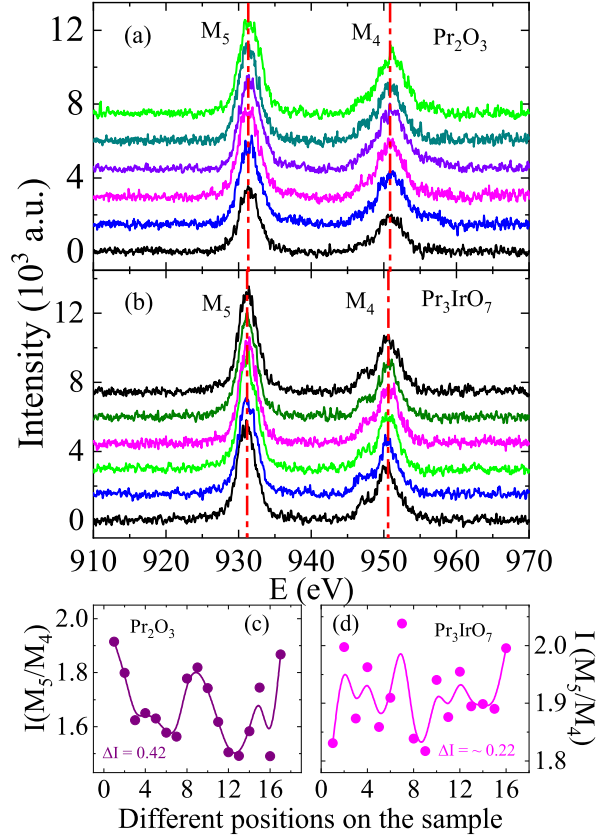}
	\caption{(a) and (b) show the EELS spectra as a function of energy for the Pr-$M_4$ and $M_5$ edges of Pr$_2$O$_3$ and Pr$_3$IrO$_7$, respectively. (c) and (d) show the relative intensity ratios of the Pr $M_5$ and $M_4$ edges at different positions of the Pr$_2$O$_3$ and Pr$_3$IrO$_7$ samples, respectively.}\label{fig.valence}
\end{figure}

In this class of oxide materials, understanding the valence state of rare-earth/transition metals is very important, since they play an important role governing the electronic, magnetic, and optical properties. As precursors, Pr$_6$O$_{11}$ and IrO$_2$ are used for the sample preparation. In Pr$_6$O$_{11}$, Pr exists in a mixed valence state of 3+/4+, i.e., 3+ for Pr$_2$O$_3$ and 4+ PrO$_2$ \cite{harish.2021}. In order to understand the Pr cationic charge state in Pr$_3$IrO$_7$, electron energy loss spectroscopy (EELS) measurements have been performed on fine powder of Pr$_3$IrO$_7$ single crystals. The Pr$_2$O$_3$ material with phase purity of $>$ 99.9\% was used as reference sample for the Pr 3+ charge state. The low loss spectra of the Dual EELS have been used for precise calibration of the zero-point. To process the EELS data, a power-law background fitting has been used. Furthermore, the spectra have been recorded as line scans for five different regions of each sample and analyzed. Fig.\ \ref{fig.valence} (a) and (b) show the Pr $M_4$ and $M_5$ edge absorption spectra for Pr$_2$O$_3$ and Pr$_3$IrO$_7$, respectively.

The $M_{4,5}$ white lines of Pr correspond to the electronic transitions between the initial state 3$d$ and the unoccupied 4$f$ orbitals. For Pr$_2$O$_3$, the $M_{5,4}$ white lines are observed around 931.3 eV and 950.8 eV, respectively. A shoulder is present on the left side of the Pr $M_4$ edge, which is indicative for the Pr 3+ valence state \cite{harish.2021,Richter.2008,Lopez.2003,Jonathan.2012}. As evident in Fig.\ \ref{fig.valence}, the line shape and peak position of the $M_{5,4}$-edge for Pr$_3$IrO$_7$ are similar to the $M_{5,4}$-edge spectra of  Pr$_2$O$_3$. A small chemical shift in the Pr $M_5$ and $M_4$ edge is observed, which amounts to $\sim$0.16~eV and $\sim$0.23~eV, resp., toward the low energy side, however, the shift is small. A shoulder is also found for Pr$_3$IrO$_7$ on the left side of the Pr $M_4$ edge. For the Pr 4+ valence state, the two peaks have been previously reported on the right side of both Pr $M_5$ and $M_4$ edges around $\sim$935~eV and $\sim$955~eV. These are distinctly absent for the present Pr$_3$IrO$_7$ sample, which also suggests that Pr is in a 3+ valence state \cite{Richter.2008, Lopez.2003}. Furthermore, we have analyzed all the other recorded regions of each sample (not shown). We have found similar types of signatures i.e., a similar amount of chemical shift in Pr $M_{5,4}$ edges, the absence of two peaks on the right side of both Pr $M_5$ and $M_4$ edges, and the presence of a shoulder on the left side of Pr $M_4$ edge. Additionally, we have further looked at the intensity ratio $I_{M_5}/I_{M_4}$ for both Pr$_2$O$_3$ and Pr$_3$IrO$_7$ materials in Fig.\ \ref{fig.valence} (c) and (d), respectively. The intensity ratio for Pr$_2$O$_3$ shows no systematic variation at different position of the sample, and the change in $I_{M_5}/I_{M_4}$ lies between 1.5 to 2 according to Fig.\ \ref{fig.valence}(c). For Pr$_3$IrO$_7$, we observe a complete fluctuation in $I_{M_5}/I_{M_4}$ ratio [see Fig.\ \ref{fig.valence} (d)], however, these fluctuations lie in the same range as for Pr$_2$O$_3$. Similar types of variation have been found in $I_{M_5}/I_{M_4}$ for other regions of the Pr$_3$IrO$_7$ sample (not shown). These results imply that Pr is in 3+ valence state, which points to a 5+ charge state of Ir.

\subsection {Magnetization study}
\begin{figure}
	\centering
	\includegraphics[width=8.5cm]{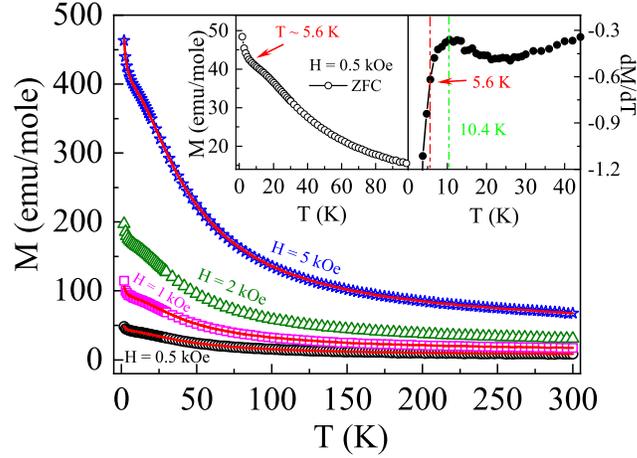}
	\caption{Temperature dependence of the dc magnetization following a ZFC and FC protocol at magnetic fields of 0.5, 1, 2 and 5 kOe for Pr$_3$IrO$_7$. The open symbols and line represent the ZFC and FC magnetic curves, respectively. Left inset: $M_{ZFC}(T)$ data measured at 0.5 kOe, where the red arrow marks the temperature 5.6 K. Right inset: Temperature derivative $dM/dT$ at a magnetic field of 0.5 kOe.} \label{fig.magnetization}
\end{figure}

Fig.\ \ref{fig.magnetization} shows the temperature-dependent dc magnetization $M(T)$ data for Pr$_3$IrO$_7$, measured for the magnetic field orientation $\textbf{H} \parallel \textbf{c}$ at 0.5, 1, 2, and 5 kOe magnetic fields by following zero field-cooled (ZFC) and field-cooled (FC) protocols. As evident in the Fig.\ \ref{fig.magnetization}, the magnetic moment in the field of 0.5 kOe increases with decreasing temperature and exhibits a broad (weak) hump in the particular temperature range of $\sim$ 22 K to $\sim$ 5.6 K and centered around $T^{*}$ $\sim$ 10.4~K. Upon further cooling the magnetization increases continuously from 5.6 K to the lowest measured temperature. For a spin-glass behavior, the magnetization is expected to exhibit a bifurcation between $M_{ZFC}$ and $M_{FC}$ and an anomaly in $M_{ZFC}$ across the glassy temperature, which shifts to lower temperature with increasing the field \cite{shobha.2021}. Our $M(T)$ does not show any such bifurcation between $M_{ZFC}$ and $M_{FC}$ in the whole temperature range, which rules out the possibility of a glassy phase. The left inset of Fig.\ \ref{fig.magnetization} depicts the low-temperature ZFC magnetization data for H = 0.5 kOe, which clearly shows the presence of a hump around $\sim$10.4~K and a prominent change around $\sim$5.6~K, as highlighted by the arrow. The $M(T)$ data measured at higher magnetic fields (1, 2, and 5 kOe) reveal that the weak hump becomes more prominent with increasing magnetic field (Fig.\ \ref{fig.magnetization}). As evident in the high field $M(T)$ data, with increasing the field the moment below $\sim$5.6~K increases continuously and exhibits a clear change in curvature of $M(T)$ in that particular temperature regime.

\begin{figure}
	\centering
	\includegraphics[width=9cm]{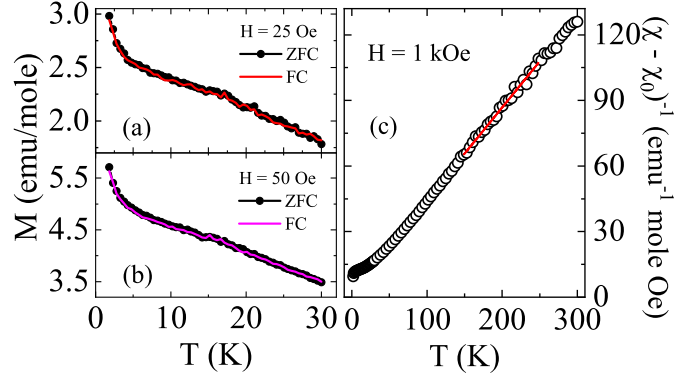}
	\caption{DC magnetization data as a function of temperature for Pr$_3$IrO$_7$ measured in low magnetic fields (a) 25~Oe and (b) 50~Oe using the ZFC and FC protocol. (c) Temperature-dependent inverse susceptibility $(\chi - \chi_0)^{-1}$ deduced from the $M_{ZFC}$ data for Pr$_3$IrO$_7$, where the solid red line represents a linear fitting describing a modified Curie–Weiss behavior as defined in the text.}\label{fig.fit-magnetization}
\end{figure}

For a more detailed analysis, we have plotted the first derivative of the $M_{ZFC}(T)$ magnetization (d$M$/d$T$) in the right inset of Fig.\ \ref{fig.magnetization} for 0.5~kOe field data. The d$M$/d$T$ data distinctly show a sharp drop below $\sim$5.6~K, and a clear change in the curvature of $M(T)$ at $T^{*}$ in the temperature range of 22~K to 5.6~K. This weak hump in $M(T)$ could be associated with the magnetic degree of freedom, which leads to a significant deviation in magnetic susceptibility from the Curie-Weiss behavior at low temperature [see Fig. \ref{fig.fit-magnetization} (c)].
It is interesting to note that a similar type of weak hump in magnetization data has been seen for other Pr based perovskites PrBO$_3$ (B = transition metals) at low temperature where the rare-earth spins order in an AFM arrangement \cite{anil.2021,gopal.2015,Mendi.2020,Mendi.2021}.
The continuous increase in magnetization below $T^{*}$ could be associated with a minor paramagnetic (PM) impurity phase consisting of non-interacting and isolated Pr or Ir moments. To obtain further insight into the low-temperature magnetic state of Pr$_3$IrO$_7$, we have measured $M(T)$ at low fields 25 and 50~Oe [see Figs.\ \ref{fig.fit-magnetization} (a) and (b), respectively]. The $M(T)$ data at the low field also exhibit a continuous increase in moment across the same temperature ($\sim$ 5.6~K), as observed in high field measurements, however, the weak hump around $\sim$ 10.4 K is not clearly visible. The notable observations in the $M(T)$ data occur at low field, namely, (i) no change in the temperature, where the magnetization continuously increases, (ii) increase in magnetic moment with increasing fields, and (iii) no hysteresis between $M_{ZFC}$ and $M_{FC}$.

Fig.\ \ref{fig.fit-magnetization} (c) depicts the temperature-dependent inverse magnetic susceptibility ($\chi^{-1}$ = H/M), which is deduced from the $M_{ZFC}$ data (Fig.\ \ref{fig.magnetization}). The $\chi^{-1}(T)$ data have been analyzed using a modified Curie-Weiss (CW) law in the temperature range 150 - 248 K \cite{harish.2017}, according to

\begin{eqnarray}
\chi(T) = \chi_0 + \frac{C}{(T - \theta_{CW})} \quad,
\end{eqnarray}

where $\chi_0$, $C$, and $\theta_{CW}$ are the temperature-independent magnetic susceptibility, the Curie constant, and Curie-Weiss temperature, respectively. The $(\chi - \chi_0)^{-1}$ vs T plot shows a linear behavior in the high temperature regime [see Fig.\ \ref{fig.fit-magnetization} (c)]. The fitting of $\chi(T)$ with Eq.~(1) gives $\chi_0$ = 9.5$\times$10$^{-3}$ emu/mole, $C$ = 2.36 emu K/mole, and $\theta_{CW}$ = - 5.7~K. The magnitude of $\theta_{CW}$ is close to $T^{*}$. The negative sign of $\theta_{CW}$ and its magnitude suggest AFM type interactions in Pr$_3$IrO$_7$. From the obtained Curie constant $C$, the effective paramagnetic moment ($\mu_{eff}$) is calculated and found to be 4.34 $\mu_B/f.u.$, and hence 2.5 $\mu_B$/Pr, which is low compared to the theoretically expected moment ($\mu_{eff} = g\sqrt{J(J+1)}$) value 3.58 $\mu_B$, where the Lande $g$ factor and the $J$ value for Pr$^{3+}$ (4$f^2$) electronic configuration are 4/5 and 4, respectively. The observed low moment ($\mu_{eff}$) implies that the four $d$-electrons of Ir$^{5+}$ (5$d^{4}$ electronic state) occupy completely $J_{eff}$ = 3/2 quartet state, which means $\mu_{eff}$ = 0 for the Ir$^{5+}$ state. This low moment ($\mu_{eff}$) discrepancy with expected moment has been seen in other Pr-based ternary oxide Pr$_3$MO$_7$ (M = Nb, Ta and Sb) materials \cite{Vente.2004}.

\begin{figure}
	\centering
	\includegraphics[width=8cm]{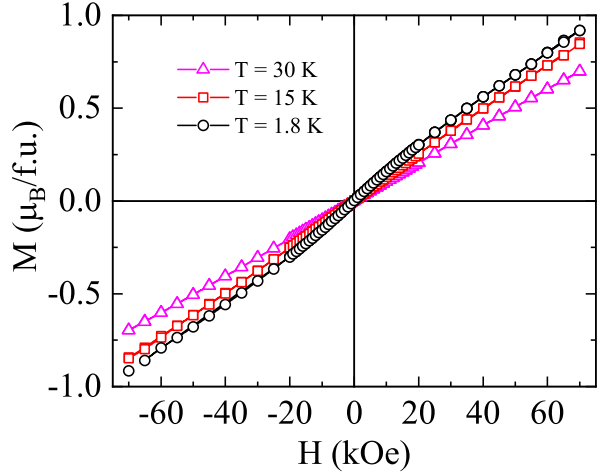}
	\caption{Magnetic field-dependent magnetization $M(H)$ collected at 1.8, 15, and 30~K.}\label{fig.field-dep-magnetization}
\end{figure}

The magnetic field-induced magnetic behavior in Pr$_3$IrO$_7$ below and above $T^{*}$ was characterized in terms of the isothermal magnetization as a function of magnetic field ($H$) for $\textbf{H} \parallel c$ at various temperatures in the field range of $\pm$ 70 kOe, and the results are shown in Fig.\ \ref{fig.field-dep-magnetization}. It is evident that the $M(H)$ data do not show any sign of saturation for all temperatures up to the maximum applied field, i.e., 70~kOe. Accordingly, the $M(H)$ data above $T^{*}$ exhibit a linear behavior like PM. In contrast, the $M(H)$ curve at 1.8 K exhibits a linear behaviour up to about $\sim$20~kOe, and above this field the $M(H)$ data show a slight deviation from the linearity with field. Furthermore, the absence of any hysteresis in the $M(H)$ curve at 1.8~K suggests that the magnetic ground state is neither a ferromagnetic nor spin-glass but could be AFM type. The magnetic moment at 1.8~K and 70~kOe is found to be 0.92 $\mu_B/f.u.$; this value is low compared to the expected moment ($\mu_H = gJ \mu_B$), which is calculated as $3.2$ $\mu_B$, where $g$ and $J$ are 4/5 and 4 for Pr$^{3+}$ (4$f^2$) electronic configuration. This low values of the magnetization and the $\mu_{eff}$ in Pr$_3$IrO$_7$ may be due to the crystal field splitting of the Pr$^{3+}$ ion because of the site symmetry of Pr(1) and Pr(2) ions \cite{Vente.2004,Zhou.2002}, which needs to be understood through microscopic probe.

\begin{figure}
	\centering
	\includegraphics[width=8cm]{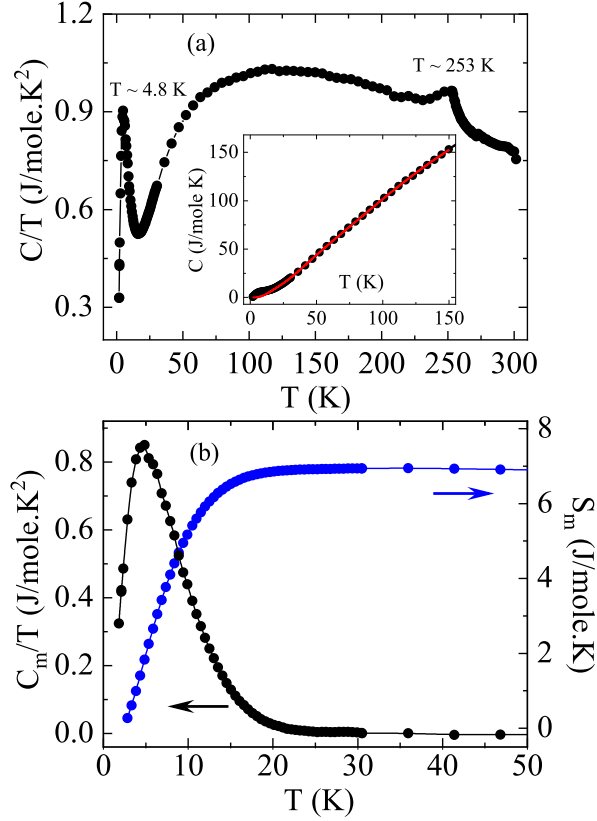}
	\caption{(a) Temperature-dependent specific heat ($C/T$) of Pr$_3$IrO$_7$. The inset shows $C(T)$ along with the lattice contributions (solid red line) obtained from fitting with the combined Debye and Einstein model. (b) Temperature-dependent magnetic specific heat ($C_m/T$) (left panel) and the calculated magnetic entropy with temperature $S_m(T)$ (right panel).
}\label{fig.heatcapacity}
\end{figure}

\subsection {Specific heat capacity}
For a better understanding of the thermodynamic behavior in the vicinity of the observed weak hump in the $M(T)$ data, the specific heat capacity $C$ as a function of temperature has been measured down to 1.8 K. Fig.\ \ref{fig.heatcapacity} shows the obtained specific heat capacity $C(T)$ in the form of a $C/T$ {\it vs} $T$ plot. The $C(T)$ data exhibit two maxima/peaks: one in the high temperature ($\sim$ 253~K) and one in the low temperature ($\sim$ 4.8~K) regime. The latter could be related to the weak hump of $M(T)$ (that manifest the maxima at $\sim$ 10.4 K), however, the small shift in temperature is not clear at the moment. A similar low-temperature change in $C(T)$ data is also present in a previous study but it has not been discussed so far \cite{Nishimine.2004}. The observed high-temperature maxima around $\sim$253 K is close to the reported value of the Pr$_3$IrO$_7$ polycrystalline sample, where it was related to a structural phase transition from orthorhombic to monoclinic \cite{Nishimine.2004, Nishimine.2007}.

Furthermore, the magnetic contribution ($C_m$) to the total heat capacity is calculated by subtracting its lattice (phonon) contribution ($C_l$) from $C$. The $C_l$ is determined by fitting of the $C(T)$ data from 30 to 150~K using a combined Debye-Einstein model ($C_l$ = $C_\mathrm{Debye}$ + $C_\mathrm{Einstein}$). The $C_\mathrm{Debye}$ and $C_\mathrm{Einstein}$ terms are given as \cite{Somnath.2017,shobha.2021}

\begin{eqnarray}
C_\mathrm{Debye} = a_{1} \left[9R \left(\frac{T}{\theta_D}\right)^3 \int^{\theta_{D}/T}_{0} \frac{x^{4} e^{x}}{\left(e^{x} - 1\right)^2} dx \right]
\end{eqnarray}

\begin{eqnarray}
C_\mathrm{Einstein} = 3R \sum^{3}_{i = 1} b_{i} \left[ \left(\frac{\theta_{Ei}}{T}\right)^2 \frac{e^{\theta_{Ei}/T}} {\left(e^{\theta_{Ei}/T }- 1\right)^2} \right]
\end{eqnarray}

where $R$ is the universal gas constant, and $\theta_D$ and $\theta_E$ are the Debye and Einstein temperature, respectively. The $a_{1}$ and $b_{i}$ coefficients are the weighting factors corresponding to acoustic and optical phonons (Debye and Einstein coefficients), respectively. The $C(T)$ data can be fitted well with one Debye and three Einstein (1D + 3E) terms of lattice specific heat and is represented by a solid line in the inset of Fig.\ \ref{fig.heatcapacity} (a). The obtained $C_l$ is extrapolated down to 1.8 K for $C_m$. From the combined fitting, the Debye and Einstein temperatures $\theta_D$, $\theta_{E_1}$, $\theta_{E_2}$ and $\theta_{E_3}$ are found to be 94.6~K, 493.7~K, 201.6~K, and 728~K, resp., and the associated weighing factors $a_{1}$, $b_{1}$, $b_{2}$, and $b_{3}$ are fixed in the ratio of 1, 5, 3, and 2, respectively. The total sum of coefficients $a_{1}$ + $\sum b_{i}$ matches well with the total number of atoms in the formula unit in Pr$_3$IrO$_7$. The obtained magnetic specific heat ($C_m$) is shown in the left panel of Fig.\ \ref{fig.heatcapacity} (b) in the form of a $C_m/T$ $vs$ $T$ plot. The magnetic specific heat shows a clear broad peak around $\sim$ 4.8 K, which is close to $T^*$ where the weak magnetic hump is observed in the $M(T)$ data.

To understand this further through the change in entropy, the magnetic entropy $S_m$ is calculated by integrating $C_m/T$ with $T$ according to
\begin{eqnarray}
\ S_m(T) = \int^{T}_{0} \frac{C_m (T)}{T} dT \quad .
\end{eqnarray}

The change in magnetic entropy with temperature $S_m(T)$  is shown in the right panel of Fig.\ \ref{fig.heatcapacity} (b). Accordingly, $S_m$ almost saturates at $\sim$ 6.9 J/mole K above $\sim$ 22.2 K, which is close to the the temperature ($\sim$ 22 K) below which the magnetization shows a change in its curvature. We have calculated the expected entropy change due to the Pr$^{3+}$ ions with $J$ = 4 and found it to be 18.2 J/(mole K) [$R$ ln(2$J$ + 1)]. The obtained $S_m$ is low and about 37.9\% of the theoretical value. This lower value of $S_m$ may be due to the higher alignment of crystal field effect levels around 22 K.

\begin{table}[b]
\caption{\label{tab:table 1} Wyckoff positions and $\Gamma$ point representations for $Cmcm$ (No. 63) space group.}
\begin{indented}
\item[]\begin{tabular}{ccc}
\hline
Atoms & Wyckoff positon & $\Gamma$ representation\\
\hline
Pr(1) &$4a$ &$ 2 B_{1u} + 2 B_{2u} + B_{3u}$\\
Pr(2) &$8g$ &$ B_{1u} + 2 B_{2u} + 2 B_{3u}$ \\
Ir &$4b$ &$ 2 B_{1u} + 2 B_{2u} + B_{3u}$\\
O(1) &$16h$ &$ 3 B_{1u} + 3 B_{2u} + 3 B_{3u}$\\
O(2) &$8g$ &$ B_{1u} + 2 B_{2u} + 2 B_{3u}$ \\
O(3) &$4c$ & $ B_{1u} + B_{2u} + B_{3u}$\\
\hline
\end{tabular}
\end{indented}
\end{table}

\begin{figure}
	\centering
	\includegraphics[width=8.5cm]{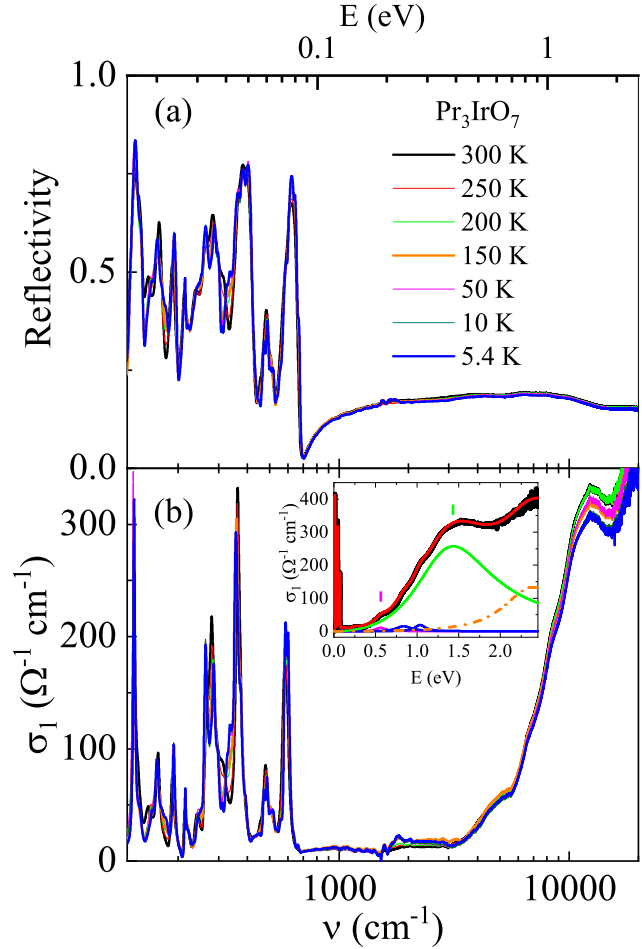}
	\caption{(a) Reflectivity and (b) optical conductivity ($\sigma_1$) of Pr$_3$IrO$_7$ at various temperatures in the whole measured frequency range. Inset in (b): Room-temperature optical conductivity $\sigma_1$ together with the Lorentz fit and the individual Lorentz contributions.}\label{fig.optics}
\end{figure}

\subsection {Optical Conductivity}
For the characterization of the electronic and vibrational properties of Pr$_3$IrO$_7$, we carried out reflectivity measurements in the infrared frequency range. The temperature-dependent reflectivity spectrum over the whole measured frequency range is shown in Fig.\ \ref{fig.optics} (a) at various temperatures, while the corresponding optical conductivity $\sigma_1$ is displayed in Fig.\ \ref{fig.optics} (b). Pr$_3$IrO$_7$ crystallises in an orthorhombic phase with $Cmcm$ symmetry. According to the factor group analysis for the $Cmcm$ structure symmetry, the irreducible representation of optical active phonons is given as

\begin{eqnarray}
        \Gamma_{optic} =  [10 B_{1u} + 12 B_{2u} + 10 B_{3u}]_{IR} + [8 A_{g} + 8 B_{1g} + 5 B_{2g} + 6 B_{3g}]_{R} \quad,
\end{eqnarray}

where $IR$ ($R$) indicates the infrared-active (Raman-active) modes. Theoretically, 32 infrared-active phonons are expected for Pr$_3$IrO$_7$ due its $Cmcm$ structure symmetry, see Table \ref{tab:table 1}. Note that all Pr, Ir and O atoms in Pr$_3$IrO$_7$ are involved in these lattice vibrations.
From the room temperature low-energy optical conductivity spectrum $\sigma_1$, two main observations can be made: First, the overall low value of $\sigma _1$ in the far-IR region, which signals the insulating character of Pr$_3$IrO$_7$. Second, 24 IR-active transverse phonon modes can be resolved and noted in Table \ref{tab:table 2}.
In analogy to the related compound Gd$_3$NbO$_7$ \cite{Ptak.2019} the modes above $\sim$190~cm$^{-1}$ can be ascribed to stretching and bending vibrations of IrO$_6$ octahedra, and the modes below $\sim$190~cm$^{-1}$ can be assigned to translational modes of the Pr$^{3+}$ and Ir$^{5+}$ ions.

The onset of electronic excitations in the optical conductivity spectrum [see inset of Fig.~\ref{fig.optics} (b)] is located at around 0.5~eV, which can serve as an estimate of the optical gap, $\Delta_{OP}$ $\sim$500~meV. The size of $\Delta_{OP}$ is comparable to that in the honeycomb-type iridates Na$_2$IrO$_3$ and $\alpha$-Li$_2$IrO$_3$ with Ir 5$d^5$ electronic configuration \cite{Comin.2012,Hermann.2017}.
The high-energy $\sigma_1$ spectrum shows weak Lorentz contributions located at $\sim$ 0.56 eV, $\sim$ 0.83 eV, and $\sim$ 1.03 eV and a stronger contribution $\sim$ 1.43~eV. In analogy to other iridates \cite{Li.2015,Kim.2016,Li.2017,Hermann.2017,Hermann.2019}, these excitations can be ascribed to on-site and inter-site transitions between $J_{eff}$ = 1/2 and $J_{eff}$ = 3/2 states, which could show additional splittings due to octahedral distortion.
The Lorentz term at $\sim$2.4~eV can be attributed to charge-transfer excitations between the Ir 5$d$ and O 2$p$ orbitals.

\begin{table}[b]
\caption{\label{tab:table 2} Observed transverse and longitudinal IR phonon mode frequencies (in cm$^{-1}$) for Pr$_3$IrO$_7$ at 300 K, 200 K, and 5.4 K.}
\begin{tabular}{cccccc}
\hline
& Transverse phonon frequencies & & & Longitudinal phonon frequencies  &\\
\hline
\hline
$300$ K & $200$ K & $5.4$ K & $300$ K & $200$ K & $5.4$ K\\
\hline
128.4 &128.4 &128.7 &132.8 &134.1 &134.4\\
133.0 &135.8 &135.3 &140.6 &140.6 &141.1\\
148.1 &151.6 &153.2 &151.2 &153.0 &154.2\\
163.6 &161.4 &161.4 &173.1 &170.2 &166.9\\
188.8 &167.5 &169.6 &195.8 &174.8 &175.4\\
193.6 &177.9 &177.2 &204.2 &178.9 &179.7\\
212.9 &189.9 &191.5 &215.3 &197.2 &198.5\\
218.6 &202.5 &201.7 &219.5 &203.4 &204.2\\
240.3 &213.6 &214.3 &242.9 &215.5 &216.5\\
265.5 &218.4 &221.4 &269.4 &221.4 &222.5\\
279.9 &229.8 &232.0 &291.4 &231.2 &233.3\\
295.5 &240.2 &243.4 &302.4 &244.3 &245.0\\
314.1 &263.6 &263.6 &325.1 &271.0 &270.5\\
362.7 &283.1 &284.8 &429.3 &293.1 &293.4\\
366.1 &297.2 &297.1 &451.4 &305.1 &304.6\\
391.5 &306.7 &311.0 &468.0 &325.6 &327.5\\
454.2 &323.2 &324.7 &495.5 &339.7 &340.0\\
478.9 &337.1 &335.6 &519.2 &391.9 &391.0\\
506.7 &358.4 &357.3 &541.0 &431.5 &426.0\\
545.0 &373.6 &373.4 &591.0 &455.8 &451.1\\
564.2 &394.7 &394.1 &633.0 &472.0 &477.1\\
579.8 &455.1 &454.1 &657.0 &495.2 &495.2\\
595.8 &471.1 &471.7 &670.7 &508.0 &506.9\\
664.6 &483.8 &485.6 &713.7 &520.7 &524.0\\
			&504.4 &504.8 & 		 &540.6 &537.2\\
			&517.3 &517.1 &			 &557.0 &556.6\\
			&542.0 &536.6 &			 &590.5 &593.5\\
			&569.9 &554.6 &			 &634.4 &637.1\\
			&585.3 &586.6 &		   &660.6 &660.3\\
			&601.2 &605.0 &			 &675.8 &679.6\\
			&628.3 &636.2 &			 &689.8 &696.8\\
			&664.7 &667.4 &			 &715.9 &717.2\\

\hline
\end{tabular}
\end{table}

\begin{figure}
	\centering
	\includegraphics[width=13cm]{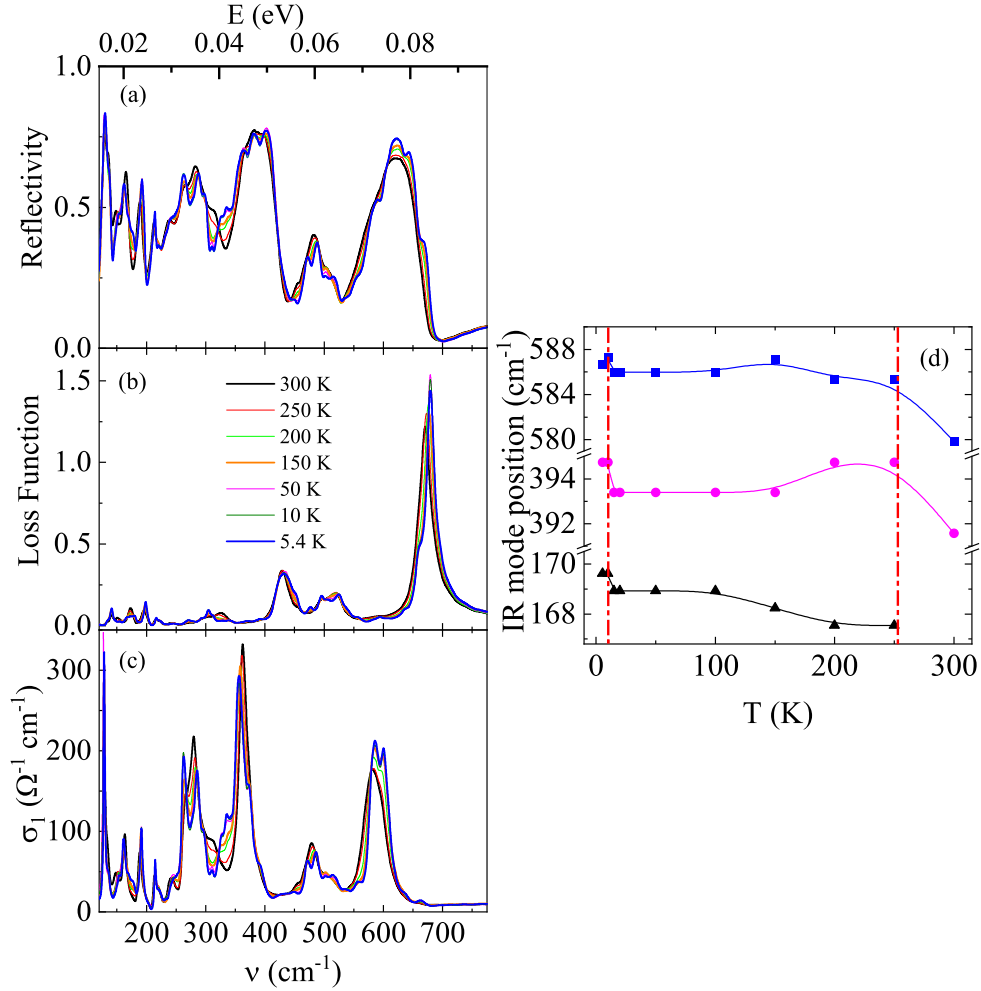}
	\caption{(a) Reflectivity, (b) loss function, and (c) optical conductivity $\sigma_1$ at various temperatures in the low-energy range. (d) Temperature evolution of the frequency of three selected phonon modes. The two vertical dashed
red lines in (d) mark the $T^*$ and structural phase transition temperature ($\sim$ 253 K).}\label{fig.Toptics}
\end{figure}

It is furthermore intriguing to compare the high-energy $\sigma _1$ spectrum of Pr$_3$IrO$_7$ with that of insulating Li$_2$RuO$_3$, where Ru$^{4+}$ has a $4d^4$ electron configuration similar to Ir$^{5+}$ (5$d^4$), however, with a smaller SOC and higher $U$-value \cite{Yun.2020}. Interestingly, for Li$_2$RuO$_3$ one observes a similar gap size $\Delta_{OP}$ $\sim$ 600 meV, and the excitations around $\sim$ 0.95 eV and $\sim$ 1.46 eV are slightly blueshifted in energy \cite{Yun.2020}, which might be due to the different SOC strength. In addition, it has been seen in a local probe study that the Ir$^{5+}$ systems exhibit higher angular strength of the SOC, $<\textbf{L$\cdot$S}>$, compared to the Ir$^{4+}$ state \cite{Laguna.2015}. Therefore, in the context of SOC, more investigations employing theoretical calculations are required to understand the electronic excitations in the strong SOC limit in Ir$^{5+}$-based oxides.

Furthermore, we checked the possible impact of the structural phase transition at $\sim$ 253 K and the magnetic ordering at $T^*$ on the vibrational and electronic excitations.
The temperature-dependent reflectivity spectrum of Pr$_3$IrO$_7$ is shown in Fig. \ \ref{fig.Toptics} (a) at various temperatures in the far-infrared range, since here the pronounced changes occur.
With decreasing temperature, we observe several changes, namely the appearance of eight additional transverse IR modes at 167.5 (20.7), 177.9 (22.0), 229.8 (28.5), 306.7 (38.0), 337.1 (41.7), 471.1 (58.4), 517.3 (64.1), 628.3 (77.8) cm$^{-1}$ (meV) around 250 K in the far-infrared range (Table \ref{tab:table 2}), a large change in the observed modes in the frequency range 290 - 350 cm$^{-1}$ (especially for the 314.1 cm$^{-1}$ mode), and the IR modes become sharper at low temperature. These changes are also clearly seen in temperature-dependent conductivity $\sigma_1$ spectrum of the far-infrared regime as shown in Fig.\ \ref{fig.Toptics} (c). The observed additional modes around 250 K are probably due to the structural phase transition which was previously reported \cite{Nishimine.2004, Nishimine.2007} and also seen in the above-described heat capacity data [see Fig.\ \ref{fig.heatcapacity} (a)].
Fig. \ref{fig.Toptics} (b) shows temperature-dependent loss function in the far-infrared range. The eigenfrequencies of the longitudinal optical phonon modes appear as maxima in the loss function \cite{Klingshirn.2012}.
We found 24 longitudinal phonon modes at room temperature and eight extra modes below the structural phase transition temperature $\sim$ 253 K, see Table \ref{tab:table 2}, which further confirms the observed extra phonon modes as seen in the optical conductivity spectrum. Like the reflectivity and conductivity spectrum, the temperature-dependent loss function exhibits a large change in the longitudinal modes in the frequency range 285 - 350 cm$^{-1}$. Moreover, an additional phonon mode around $\sim$ 371.0 cm$^{-1}$ is seen in the loss function spectrum compared to conductivity spectrum.
The temperature-dependent $\sigma_1$ spectrum in the higher energy region is shown in the Fig. \ \ref{fig.optics} (b). Obviously, the observed Lorentz contributions ($\sim$ 0.56 eV, $\sim$ 0.83 eV, $\sim$ 1.03 and $\sim$ 1.43 eV) do not show any noticeable change with lowering the temperature. Fig. \ \ref{fig.Toptics} (d) shows the frequency position of three selected phonon modes as a function of temperature. The frequency of the modes at 579.8 and 391.5 cm$^{-1}$ show an abrupt increase around 250 K, which we attribute to the structural phase transition occurring at this temperature.
All three selected modes show an anomaly at around $T^*$, where a weak hump is observed in the magnetization data.
It is well known that the Dzyaloshinskii-Moriya interaction (DMI) mainly occurs due to the distortions of IrO$_6$ octahedra which are induced by the presence of strong SOC in iridates \cite{Bhatti.2014}. In recent studies, it has been shown that spin-phonon coupling occurs due to the bending of Ir-O-Ir bonds which is mainly mediated by the DMI \cite{jae.2019, harish.2019}. For Ln$_3$IrO$_7$ (Ln = Pr, Nd, Sm and Eu), it has been reported that both tilt of the IrO$_6$ octahedra and distortion in the Ir-Ln-O slab increase at the low temperature \cite{Nishimine.2007}. Therefore, the observed an anomaly for the IR phonon modes around $T^*$ underlines the importance of spin-phonon coupling in Pr$_3$IrO$_7$.

\section{Conclusion}
In summary, the structural, magnetic, thermodynamic, and optical properties are studied in detail for single-crystalline Pr$_3$IrO$_7$ ternary iridate. The structural analysis shows that Pr$_3$IrO$_7$ crystallizes in the orthorhombic phase with $Cmcm$ symmetry. The electron energy loss spectroscopy measurements reveal a 3+ oxidation state for Pr, hence 5+ oxidation state of Ir in Pr$_3$IrO$_7$. As a consequence, it  represents the ideal system to study the magnetic and electronic evolution of Pr rare-earth. Our magnetic and specific heat results suggest a possible weak low-temperature AFM state for Pr$_3$IrO$_7$. Based on the reflectivity data analysis, we conclude an insulating electronic structure for Pr$_3$IrO$_7$ with an optical gap $\Delta_{OP}$$\sim$500~meV. The temperature-dependent reflectivity spectrum shows eight extra phonon modes below 250 K, which can be attributed to a structural phase transition. Furthermore, several phonon modes reveal an anomaly in the temperature dependence of their frequency position at $T^*$ $\sim$ 10.4 K, where a weak hump is observed in the magnetization data, indicative for sizeable spin-phonon coupling in the system.

\ack{H. K. acknowledges Dr.\ J. Ebad-Allah and Dr.\ S. Ghara for fruitful discussions. H. K. acknowledges funding (PDOK-95-19) from the Bavarian Science Foundation and by the Deutsche Forschungsgemeinschaft, Germany, through Grant No.\ KU 4080/2-1.}

\end{document}